\documentclass[a4paper,12pt]{article}

\usepackage{jheppub}
\usepackage[utf8]{inputenc}
\usepackage[english]{babel}

\usepackage{amsfonts}
\usepackage{amsthm}
\usepackage{mathrsfs}
\usepackage[mathcal]{euscript}
\usepackage{float}
\usepackage{latexsym}
\usepackage{subfig}

\newcommand{\be}{\begin{equation}}
\newcommand{\ee}{\end{equation}}

\newcommand{\skipline}{\vspace{\baselineskip}}

\usepackage{mathtools}
\DeclarePairedDelimiter\corrfunc{\langle}{\rangle}

\newcommand{\nn}{\nonumber}
\newcommand{\eps}{\varepsilon}
\newcommand{\om}{\omega}
\newcommand{\largemom}{\underset{k\,\to\,\infty}{\approx}}

\newcommand{\largetimesim}{\underset{t\,\to\,\infty}{\sim}}
\newcommand{\Z}[1]{\mathbb{Z}_{#1}}
\newcommand{\figref}[1]{Fig.~\ref{#1}}
\newcommand{\secref}[1]{section~\ref{#1}}

\usepackage{comment}
\usepackage{xcolor}
\usepackage{soul}

\definecolor{dgreen}{rgb}{0,0.6,0}

\definecolor{brown}{rgb}{0.56,0.22,0.22}

\title{Quantum quenches in fractonic field theories}

\author{Dmitry S. Ageev}
\author{and Vasilii V. Pushkarev}

\affiliation{Steklov Mathematical Institute, Russian Academy of Sciences,\\
Gubkin str. 8, 119991 Moscow, Russian Federation}

\emailAdd{ageev@mi-ras.ru}
\emailAdd{pushkarev@mi-ras.ru}

\abstract{We study out-of-equilibrium dynamics caused by global quantum quenches in fractonic scalar field theories. We consider several types of quenches, in particular, the mass quench in theories with different types of discrete rotational symmetries ($\Z4$ and $\Z8$), as well as an instantaneous quench via the transition between them. We also investigate fractonic boundary quenches, where the initial state is prepared on a finite-width slab in Euclidean time. We find that perturbing a fractonic system in finite volume especially highlights the restricted mobility via the formation and subsequent evolution of specific $\Z4$-symmetric spatial structures. We discuss a generalization to $\Z{n}$-symmetric field theories, and introduce a proper regularization, which allows us to explicitly deal with divergences inherent to fractonic field theories.}

\begin{document} 
\maketitle
\flushbottom

\section{Introduction}

The study of quantum systems that are taken out of equilibrium possesses a great challenge for modern physics, being at the same time a key ingredient for understanding many fundamental aspects of quantum field theory and providing promising prospects for applications.

Any external force acting fast enough inevitably takes an isolated system out of equilibrium. It does not matter what type of force it is and how it is described if we are going to study the correlation-function dynamics induced by excitation. In many cases, the entire process of excitation can be viewed as an instant change of model parameters or as the preparation of an initial state which is not an eigenstate of the Hamiltonian. A range of models based on this idea have been studied, generally being referred to as quantum quenches. The elementary examples of quantum field theories, such as conformal field theories (CFTs)~\cite{Calabrese:2006rx, Calabrese:2007rg, Das:2014hqa, Calabrese:2016xau}, free field theories~\cite{Sotiriadis2009thermal, Rajabpour:2014osa, Das:2015jka, Das:2016lla} and simplest interacting theories~\cite{Sotiriadis:2010si} have been explored in the past two decades. More advanced cases, including non-conformal operator local quench~\cite{He:2019vzf, Ageev:2022kpm}, have been actively studied recently. Theoretical investigation of quenches proved to be the most intensive and successful in the holographic framework~\cite{Danielsson:1999fa, Balasubramanian:2010ce, Balasubramanian:2011ur, Buchel:2013gba, Asplund:2014coa, Ageev:2017oku, Ageev:2017wet} due to potential connections with black hole formation. Also, the problem of quantum quenches gives rise to fruitful results for understanding the evolution of entanglement entropy~\cite{Calabrese:2005in, Nozaki:2014hna, He:2014mwa, Caputa:2014eta, Cotler:2016acd, Mozaffar:2021nex}.

One of the most fundamental concepts of quantum field theory is symmetries. A particular symmetry possessed by a system defines the dynamical features of observables. Hence, the change of symmetry should lead to a subsequent change in the dynamics. Since a Lagrangian generically has a fixed set of symmetries, switching between different sets also changes the Hamiltonian, so it can be viewed as a process preparing an out-of-equilibrium state of the system. Despite the fact that most papers on quenches in quantum field theory have been restricted mainly to relativistic theories, i.e., invariant under the Lorentz group rotations, some examples of long-range field theories, which are not Lorentz-invariant, have been studied. They include theories with general power law-like dispersion relation~\cite{Rajabpour:2014osa, Mozaffar:2021nex} and Lifshitz harmonic model~\cite{MohammadiMozaffar:2018vmk, MohammadiMozaffar:2019gpn}. It is expected that non-relativistic setups are the most phenomenologically interesting for describing phases of matter in condensed matter theory.

In this paper, we study global quenches in fractonic field theories. Recently discovered fractonic quantum systems are among examples of theories with exotic global symmetries~\cite{Vijay:2015mka, Pretko:2018jbi, Gromov2019towards, Radzihovsky:2019jdo, Seiberg:2019vrp, Seiberg:2020bhn, Seiberg:2020wsg, Distler:2021bop, Bulmash:2023msp}, which possess in their spectra quasi-particles of restricted mobility, called \emph{fractons}.\footnote{Interesting to note but the term \emph{fractons} was introduced with different meaning long before,~\cite{Khlopov1981fractionally}, for denoting particles with fractional charges.} In this work, by the fractonic quantum field theory we mean a low-energy continuum limit of $2 + 1$-dimensional XY-plaquette lattice model, in which scalar field interactions take place around plaquettes,~\cite{Paramekanti2002ring, Seiberg:2020bhn}. The theory has the momentum dipole global symmetry,
\be\nn
    \phi(t, x, y) \to \phi(t, x, y) + g_x(x) + g_y(y),
\ee
for arbitrary functions~$g_x$ and~$g_y$, as well as $\Z4$ discrete rotational symmetry. Fractons, being immobile quasiparticle defects, contrast in this regard with the particles of the relativistic quantum field theory. The other significant property is that the ground state of the fractonic theory is highly degenerate. 

The very first subtle point concerning the study of fractonic post-quench dynamics is the UV/IR mixing, which is a consequence of the ground-state degeneracy. This intrinsic feature of fractonic theories leads to severe divergences of correlation functions. Hence, one has to impose a regularization both on UV and IR degrees of freedom to get reasonable insights on the dynamical picture.\footnote{In~\cite{Distler:2021bop}, the regularization by rotationally invariant momentum cut-off was exploited for beta-function calculations.} For this purpose, we introduce double regularization consisting of turning on a mass gap and small \emph{relativistic regularization} by the inclusion of a relativistic term in the theory. These terms explicitly break the momentum dipole symmetry in a controllable way, i.e., making them smaller brings the theory closer to the purely fractonic case and hence one can observe fractonic dynamical effects. We find out that this regularization tames the divergences in the stationary case. However, it turns out that the study of non-equilibrium dynamics requires additional treatment of new divergent pieces, and we propose how to do that.

Adding the mass gap and the relativistic regularization break the momentum dipole symmetry, while preserving discrete rotational symmetry~$\Z4$. We introduce the generalization of the action to discrete rotational symmetries of higher orders. In the momentum space, this corresponds to a change in the invariant under the symmetry group momentum polynomials included in the action.\footnote{During the preparation of this work, we became aware of the paper, which considers multipole symmetries for an arbitrary group order in fractonic theories in detail, see~\cite{Bulmash:2023msp}.}

As the starting point, we use the model of a global quantum quench described in the series of works~\cite{Calabrese:2007rg, Sotiriadis2009thermal, Sotiriadis:2010si}. The instant global change of the mass gap in scalar free field theory takes the system out of equilibrium leading to a different dispersion relation. In the case of a Gaussian field theory, the effect of the quench can be traced out analytically. In this paper, we implement the model of the global quench to the study of fractonic field theories. In addition to the change of mass, it is natural to consider, as a setup for a global quench, an instant switch from the Lorentz symmetry to the fractonic symmetry, while keeping other parameters the same. One can consider this as turning on highly constrained dynamics. Another setup is the so-called boundary quench, in which an out-of-equilibrium state is prepared as a path integral on a finite-width slab in Euclidean time with boundary conditions imposed on the slab boundaries. We also study the dynamics after such a quench in the case of finite volume and find that the restricted mobility manifests itself in a clear manner through the formation of complicated structures during time evolution.\footnote{We refer the reader to recent paper~\cite{Islam:2023cmm}, in which the phase diagram in the context of the quench in fractonic field theories has been discussed.}

The paper is organised as follows. In \secref{sec:setup}, we describe the particular quantum field theory models that we apply calculations for. The quench, in which dispersion relation is instantly changed meaning that the symmetry of the problem is changed as well, is considered in \secref{sec:mass-quench}. In \secref{sec:slab-quench}, the dynamics following the boundary quench is described. In \secref{sec:finvolume}, we study the same theory being putted into finite volume. We end with outlook in \secref{sec:conclusions}.

\section{Field theory model: regularization and \texorpdfstring{$\Z{n}$}{Zn} symmetries}
\label{sec:setup}

In this paper, we focus on the $2 + 1$-dimensional free scalar field theory described by the following Euclidean action
\be 
    S = \frac{1}{2}\int d\tau\,dx\,dy\,\phi \hat{D} \phi,
\ee 
with spatial coordinates $x,y$ and Euclidean time $\tau$. The differential operator $\hat{D}$ in the momentum space is given by
\be 
    \hat{D} = \om^2 + \eps\left(k_x^2 + k_y^2\right) + f(k_x, k_y) + m^2,
\ee 
with the momenta $k_{x,y}$ corresponding to the spatial coordinates $x$ and $y$, respectively. Such a choice of $\hat{D}$ in free field theory yields the dispersion relation of the form
\be 
    \om_k = \sqrt{\eps\left(k_x^2 + k_y^2\right) + f(k_x, k_y) + m^2},
\ee 
where $f(k_x, k_y)$ is a function to be fixed further, which reflects a particular discrete symmetry of a fractonic field theory under consideration. We also assume that the general form of the dispersion relation includes a rotationally invariant part $\eps(k_x^2 + k_y^2)$, which is necessary to regulate the divergences corresponding to the UV/IR mixing.

One should notice the emergence of the highly degenerate ground state in this model~\cite{Seiberg:2020bhn}, thus, expecting the presence of UV/IR mixing and divergences even in the finite-volume case. The inclusion of the mass gap $m$ fixes a divergence arising at $k_x = k_y = 0$. The relativistic regulator eliminates the divergences corresponding to large momentum values along one of the axes with zero value along the other axis, i.e., when $k_x = 0, k_y \to \infty$ or $k_y = 0, k_x \to \infty$. The introduction of the regulators breaks the momentum dipole symmetry in a controllable way: making them small enough we study ``nearly-fractonic field theories''. The regularization makes it possible to study emergent fractonic effects of out-of-equilibrium dynamics by tuning~$\eps$ and~$m$. At the same time, it is interesting to observe how the presence of the mass gap as well as the interplay between the fractonic and the relativistic parts of the kinetic term affect out-of-equilibrium properties of the model on its own.

\skipline

Thus, we assume that the dispersion relation of the \emph{regularized} fractonic model with $\Z4$ symmetry is given by
\be 
    \Z4\text{-symmetric model:} \quad \om_k = \sqrt{\eps\left(k_x^2 + k_y^2\right) + k_x^2 k_y^2 + m^2}.
    \label{eq:Z4disprel}
\ee

One of the main examples of fractonic field theories, a low-energy limit of the so-called XY-plaquette model~\cite{Seiberg:2020bhn}, corresponds to the choice $\eps = 0$, zero mass $m = 0$, and the special form of the function $f(k_x, k_y)$ such that $f_{XY}(k_x, k_y) = k_x^2 k_y^2$. In the coordinate space, this setup is defined by a higher-derivative action
\be
    S = \frac{1}{2}\int d\tau\,dx\,dy \left((\partial_{\tau}\phi)^2 + (\partial_x\partial_y\phi)^2\right).
    \label{eq:Fracton_action}
\ee
The trivial choice $f(k_x, k_y) = 0$ with $\eps = 1$ leads to the relativistic massive free field theory with the standard dispersion relation
\be 
    \text{Relativistic model:} \quad \om_k = \sqrt{k_x^2 + k_y^2 + m^2}.
    \label{eq:reldisprel}
\ee

\skipline

The action~\eqref{eq:Fracton_action} is an example of a fractonic field theory with $\Z4$ symmetry. Higher order discrete symmetries allow for different versions of the action due to the growth of the number of polynomials invariant under the symmetry group. The $\Z4$-symmetric choice of the function $f(k_x, k_y)$ after the transformation to polar variables in the momentum space $k_x = k\cos\varphi$ and $k_y = k\sin\varphi$,
\be 
    f_{XY} = k^4 \cos^2\varphi \sin^2\varphi,
\ee 
allows for a generalization by changing the period of the angular variable,
\be
    f_\alpha = k^4 \cos^2(\alpha\varphi) \sin^2(\alpha\varphi) = -\frac{\left(-\left(k_x^2 + k_y^2\right)^{2\alpha} + \left(k_x + i k_y\right)^{4\alpha}\right)^2}{16\left(k_x^2 + k_y^2\right)^{2\alpha - 2}\left(k_x + i k_y\right)^{4\alpha}}.
    \label{eq:alphamom}
\ee
For instance, the expressions for lowest values of $\alpha$ take the form
\begin{gather} 
    f_\alpha\Big\rvert_{\alpha\,=\,2} = \frac{4\left(k_x^3 k_y - k_x k_y^3\right)^2}{\left(k_x^2 + k_y^2\right)^2}, \quad f_\alpha\Big\rvert_{\alpha\,=\,3} = \frac{\left(3 k_x^5 k_y - 10 k_x^3 k_y^3 + 3 k_x k_y^5\right)^2}{\left(k_x^2 + k_y^2\right)^4}.
\end{gather}
The function $f_\alpha$ corresponds to some non-local operators in the coordinate space. In fact, one could construct a local higher-derivative action with the operator $\hat{D}$ defined by this invariant momentum-space polynomial. The numerator of $f_\alpha$ (e.g. $(k_x^3 k_y - k_x k_y^3)^2$ for the case $\alpha = 2$) is already an invariant polynomial of order $4\alpha$ under rotations on $90/\alpha$ degrees, providing $\Z{4\alpha}$ symmetry to the action. However, we would like to keep the form~\eqref{eq:alphamom} with the denominator, which gives the same asymptotic IR structure as the $\Z4$-symmetric model, $\sim k_xk_y$. We emphasize that the polynomial obtained in this way is only an instance of invariant polynomials; the full treatment of irreducible representations of multipole groups is made in~\cite{Bulmash:2023msp}, while the general description of the construction of effective field theories is given in~\cite{Gromov2019towards}.

The $\Z8$-symmetric model (which will be the second example of a fractonic theory to consider in what follows) also requires a regularization, hence, the form of the dispersion relation suitable for practical calculations reads as
\be
    \Z8\text{-symmetric model:} \quad \omega_k = \sqrt{\eps(k_x^2 + k_y^2) + \frac{4(k_x^3 k_y - k_x k_y^3)^2}{(k_x^2 + k_y^2)^2} + m^2}.
    \label{eq:Z8disprel}
\ee

\section{Instant change of dispersion relation}
\label{sec:mass-quench}

\subsection*{Preliminaries}

A global quench takes the system out of equilibrium by an instant global excitation, which in the relativistic massive free scalar field theory can be implemented via the change of the mass gap from some initial value $m_0$ to a final value $m$ at a fixed time moment $t_0 = 0$. This can be interpreted as a change of the dispersion relation describing the system, from the initial $\om_{k,0}$, depending on the value $m_0$, to the final one $\om_k$, depending on $m$. This type of excitation can be considered in a more general form as an instantaneous change in the dispersion relation itself without taking into account the change in mass. In what follows, we consider the transition between dispersion relations of different forms corresponding to changes of Lagrangians and, hence, to changes in the behavior of the system. Let us list some of the options:
\begin{itemize}
    \item the change of theory parameters (for example, the mass), while keeping the symmetry unchanged;
    \item transition from the relativistic dispersion relation to the fractonic one;
    \item switching off the fractonic dispersion relation with transition to the ordinary relativistic behavior;
    \item transition between the instances of the fractonic symmetry.
\end{itemize}

It was derived in~\cite{Calabrese:2007rg, Sotiriadis:2010si} that the dynamics of the two-point function in free field theories after global quench with the change of dispersion relation is exactly solvable and described by the Fourier transform
\be
    \corrfunc{\phi(t_1, x, y)\phi(t_2, 0, 0)} = \int\frac{\,dk_x\,dk_y}{(2\pi)^2}e^{ik_xx + ik_yy} \mathcal{F}(t_1, t_2, k_x, k_y),
\ee
where the function $\mathcal{F}$ is given by
\be
    \begin{aligned}
        \mathcal{F}(t_1, t_2, k_x, k_y) & = \frac{(\om_k - \om_{k,0})^2}{4\om_k^2\om_{k,0}}\cos\om_k(t_1 - t_2) + \frac{\om_k^2 - \om_{k,0}^2}{4\om_k^2\om_{k,0}}\cos\om_k(t_1 + t_2) + \frac{e^{-i\om_k|t_1 - t_2|}}{2\om_k}.
    \end{aligned}
\ee
This non-equilibrium two-point correlator is translationally invariant with respect to the spatial coordinates, but the time-translational invariance is broken by the quench.

The result was obtained as a generalization of the quench of a single harmonic oscillator to the quench of a linearly coupled harmonic oscillator. Therefore, it should be valid for dispersion relations of any free scalar field theory, i.e., which Hamiltonian is diagonalizable in the momentum space. The formula can be used both in the quench by mass as well as when the underlying symmetry is instantly changed via the switch between dispersion relations. 

To gain a better visibility of the non-stationary effects, in what follows, we consider the evolution of the equal-time two-point correlation function with the initial value subtracted (as in~\cite{Calabrese:2007rg, Sotiriadis2009thermal}), namely,
\be
    \begin{aligned}
        G(t, x, y) & \equiv \corrfunc{\phi(t, x, y)\phi(t, 0, 0)} - \corrfunc{\phi(0, x, y)\phi(0, 0, 0)} = \\
        & = \int\frac{\,dk_x\,dk_y}{(2\pi)^2}e^{ik_xx + ik_yy}\mathcal{G}(t, k_x, k_y),
    \end{aligned}
    \label{eq:mass-quench}
\ee
where $\mathcal{G}$ is given by
\be
    \mathcal{G}(t, k_x, k_y) = \frac{\left(\om_{k,0}^2 - \om_k^2\right) \sin^2(\om_k t)}{2\om_k^2 \om_{k,0}}.
    \label{eq:FourierImage_mass}
\ee
The regularized two-point correlation function $G(t, x, y)$ describes the perturbations caused purely by the quench. In the case when there is no change of the dispersion relation, i.e., no quench at all, this function vanishes identically.

\begin{figure}[t]\centering
    \includegraphics[width=0.8\textwidth]{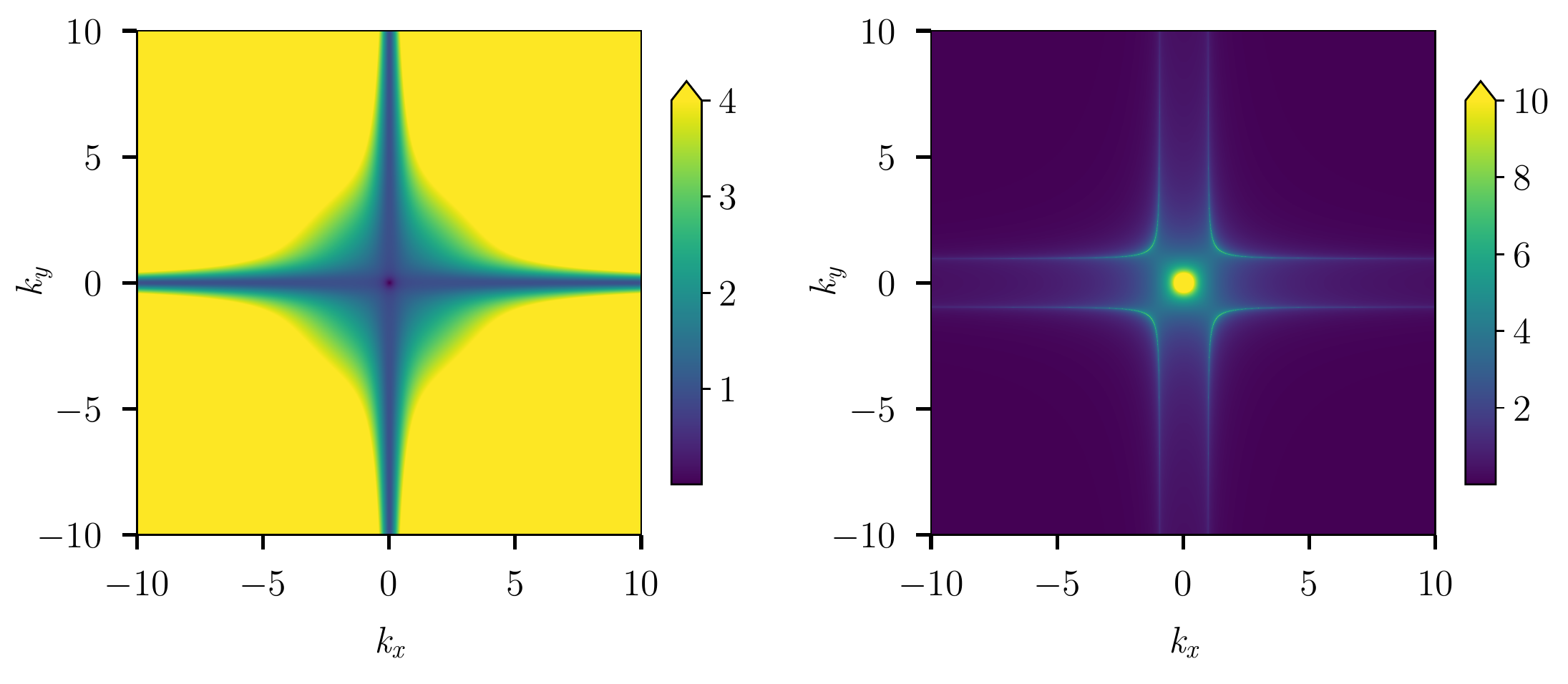}
    \caption{\textit{Left:} Group velocity~\eqref{eq:vel} for the regularized fractonic dispersion relation~\eqref{eq:Z4disprel}. The parameters are $m = 0.3$, $\eps = 1$. \textit{Right:} The dependence of the inverse effective temperature $\beta_\text{eff}$~\eqref{eq:EffTemp} on the momentum in the relativistic-to-fractonic quench; $m = 0.3$, $\eps = 0.1$. There is an infinite set of modes that equilibrate to zero temperature.}
    \label{fig:therm}
\end{figure}

\subsection*{Excitation velocity and effective temperature features}

As it was discovered in works on global quenches in CFT,~\cite{Calabrese:2007rg}, and in long-range theories,~\cite{Rajabpour:2014osa}, the post-quench dynamics is characterized by two main features in the semiclassical approximation. The first one is that there are pairs of excitations propagating with classical group velocity $v = |\nabla_\mathbf{k}\om_k|$, forming a causal region in cases when this velocity has a maximum, and the second one is that the final state is described in terms of the Generalized Gibbs Ensemble thermalization proposition.

\skipline

In the case of the relativistic dispersion relation, the existence of the maximum velocity, $v_\text{max} = 1$, leads to the emergence of the effective horizon that forbids correlations to come at a given point, distanced from the quench point by $r$, until $t = r/2$. In the case of the massive fractonic dispersion relation, oppositely, there are acausal modes that propagate with infinitely large velocity since there is no maximum of $v$,
\be
    v = \sqrt{\frac{k_x^2 + k_y^2}{1 + \displaystyle\frac{m^2}{k_x^2 k_y^2}}},
\ee
even if the relativistic regularization~\eqref{eq:Z4disprel} is introduced,
\be
    v_{\eps} = \sqrt{\frac{\eps^2\left(k_x^2 + k_y^2\right) + 4\eps k_x^2k_y^2 + k_x^2k_y^2\left(k_x^2 + k_y^2\right)}{\eps\left(k_x^2 + k_y^2\right) + k_x^2k_y^2 + m^2}},
    \label{eq:vel}
\ee
see~\figref{fig:therm}, left. Non-regularized fractonic field theory possesses an infinite set of frozen modes such that one component of~$\mathbf{k}$ equals zero, while the other is constant. The relativistic regularization by~$\eps$ enables such modes to propagate with the maximum velocity, $\max_{k_x}(\eps k_x/\sqrt{\eps k_x^2 + m^2})$, being equal to~$\sqrt{\eps}$. This is expected since the relativistic regulator turns off the degeneracy of such modes by adding relativistic degrees of freedom. The existence of the maximum velocity for such modes should lead to the formation of the wave front in the profile of the two-point correlation function propagating as fast as~$2\sqrt{\eps}$. In the $\Z8$-symmetric case~\eqref{eq:Z8disprel}, the situation is the same, except that the modes $k_x = \pm k_y$ are also frozen in the non-regularized case.

\skipline

Another aspect of quenches in free field theories is thermalization with the final distribution of modes according to the Generalized Gibbs Ensemble, meaning that each mode at large enough time after the quench equilibrates to some finite effective temperature. Since the dispersion relation~$\om_k$ has a stationary point (a local minimum) at $(k_{x,0}, k_{y,0}) = (0, 0)$, we expect that the equilibration takes place. Indeed, using the stationary phase method, it can be shown~\cite{Calabrese:2007rg} that the large-time asymptotic behavior of the integral~\eqref{eq:mass-quench} is given by
\be
    G(t, x, y) \largetimesim t^{-1}\cos\left(2mt\right),
\ee
which tends to zero value meaning that the equilibration occurs.

\begin{figure}[t]\centering
    \includegraphics[width=1.\textwidth]{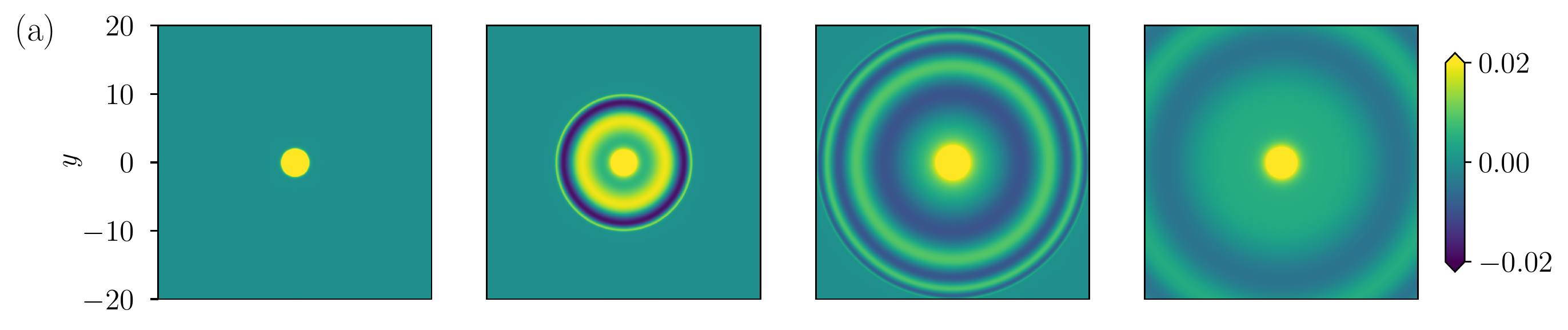}
    \includegraphics[width=1.\textwidth]{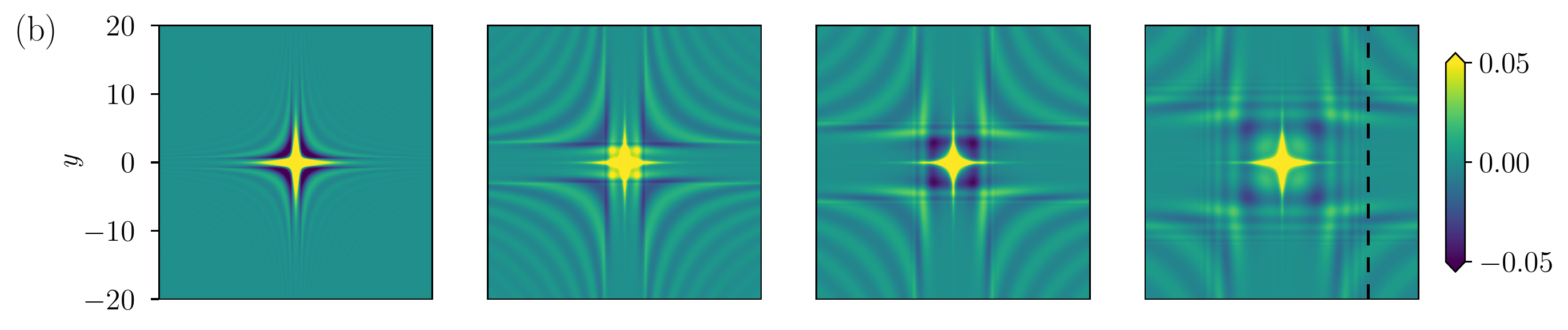}
    \includegraphics[width=1.\textwidth]{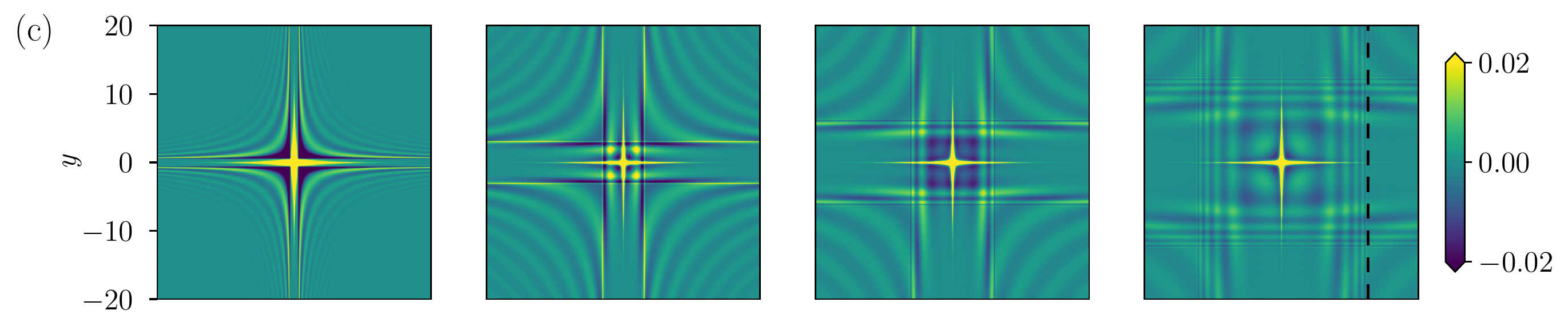}
    \includegraphics[width=1.\textwidth]{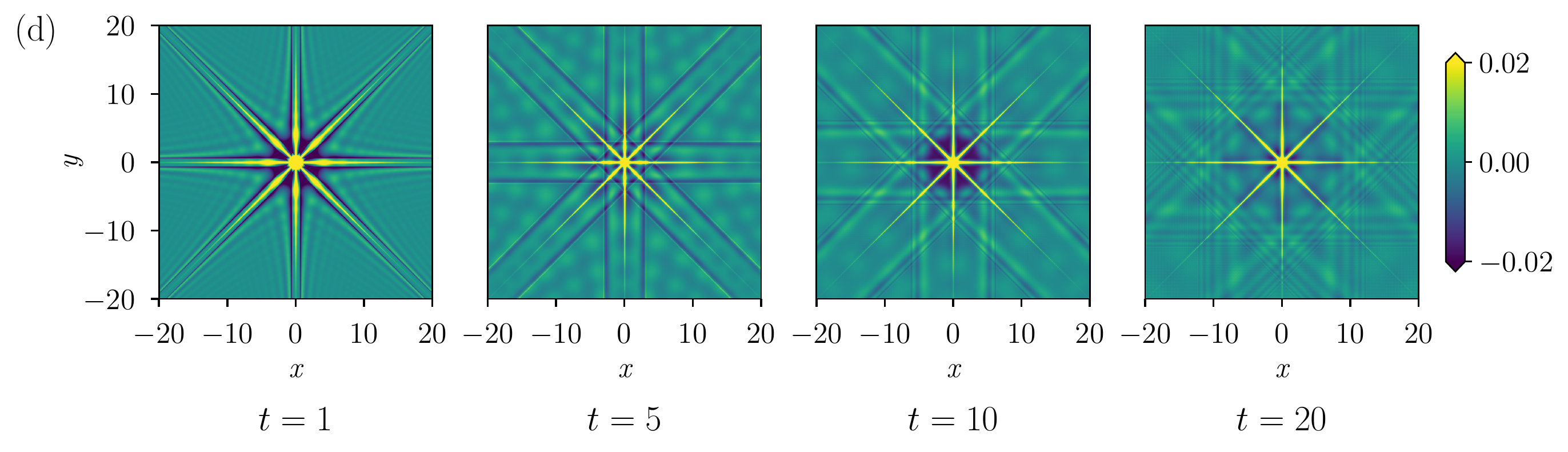}
    \caption{Regularized two-point correlation function~\eqref{eq:mass-quench} at different time moments when: \textit{(a)}~mass is changed from $m_0 = 5$ to $m = 1$ in the relativistic theory; \textit{(b)}~mass is changed from $m_0 = 5$ to $m = 1$ in the $\Z4$-symmetric fractonic theory, $\eps = 0.1$; \textit{(c)}~relativistic symmetry is changed to $\Z4$ one, $m = 1$, $\eps = 0.1$; \textit{(d)}~relativistic symmetry is changed to $\Z8$ one, $m = 0.3$, $\eps = 0.1$. Black dashed lines mark a position $2\sqrt{\eps}t$ (analytical expression for the position of the wave front).}
    \label{fig:DispChangeEvolPart1}
\end{figure}

In~\cite{Calabrese:2007rg}, it was also shown that the inverse effective temperature~$\beta_\text{eff}$ for each mode is momentum-dependent, and for a generic dispersion relation of a free theory is given by
\be
    \beta_\text{eff}(k_x, k_y) = \frac{2}{\om_k}\ln\frac{\om_k + \om_{k,0}}{|\om_k - \om_{k,0}|}.
    \label{eq:EffTemp}
\ee
There are three cases, which can be distinguished in the behavior of this function:
\begin{itemize}
    \item the transition into the same symmetry class (i.e. mass quench) is characterized by the finite inverse effective temperature for each mode;
    
    \item the transition of the form $\Z4 \to \Z8$ (or vice versa) with mass being fixed possesses a divergence of the inverse effective temperature along the lines $k_x = 0$ and $k_y = 0$;
    
    \item the inverse effective temperature corresponding to the relativistic-to-fractonic (or vice versa) quench with mass being fixed is logarithmically divergent at the vanishingly small absolute value of the momentum, as well as on the surface $(1 - \eps)(k_x^2 + k_y^2) - k_x^2k_y^2 = 0$, see~\figref{fig:therm}, right. Therefore, there is an infinite set of modes equilibrating to zero effective temperature ($\beta_\text{eff} \to \infty$),
        \be
            (k_x, k_y) = \left\{
            \begin{aligned}
                & \left(k_x, \pm\frac{k_x\sqrt{1 - \eps}}{\sqrt{\eps - 1 + k_x^2}}\right), \quad \eps \neq 1, \\
                & \left(0, k_y\right), k_y \in (-\infty, \infty) \quad\text{or}\quad \left(k_x, 0\right), k_x \in (-\infty, \infty), \quad \eps = 1.
            \end{aligned}\right.
        \ee
\end{itemize}

\subsection*{Evolution picture}

\begin{figure}[t]\centering
    \includegraphics[width=1.\textwidth]{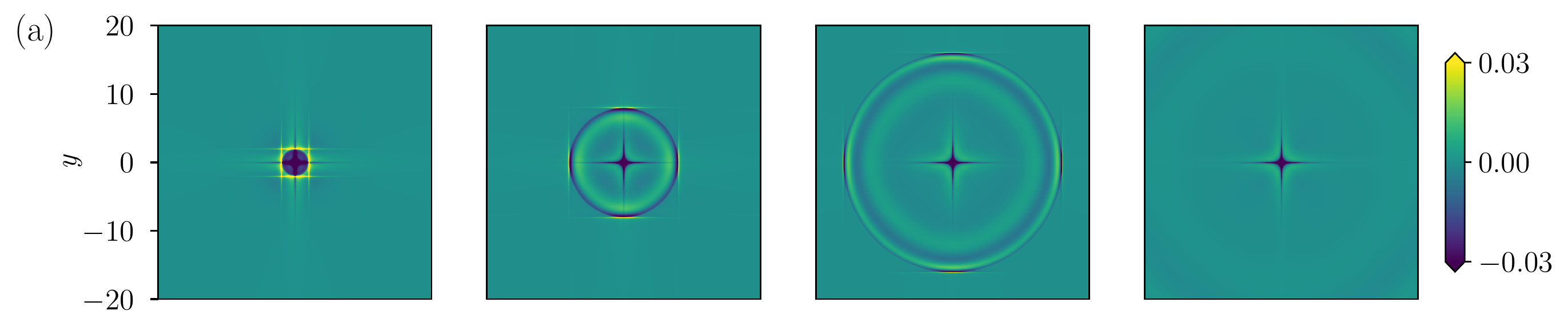}
    \includegraphics[width=1.\textwidth]{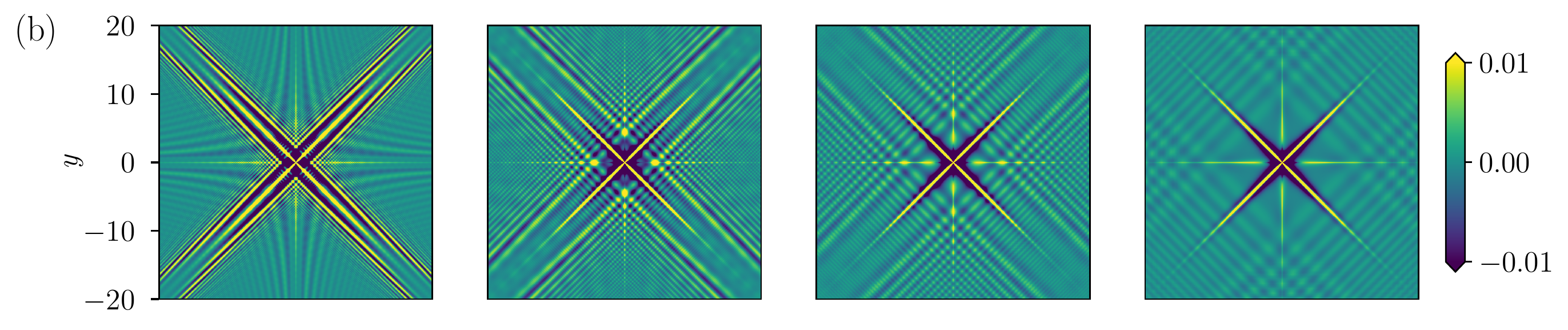}
    \includegraphics[width=1.\textwidth]{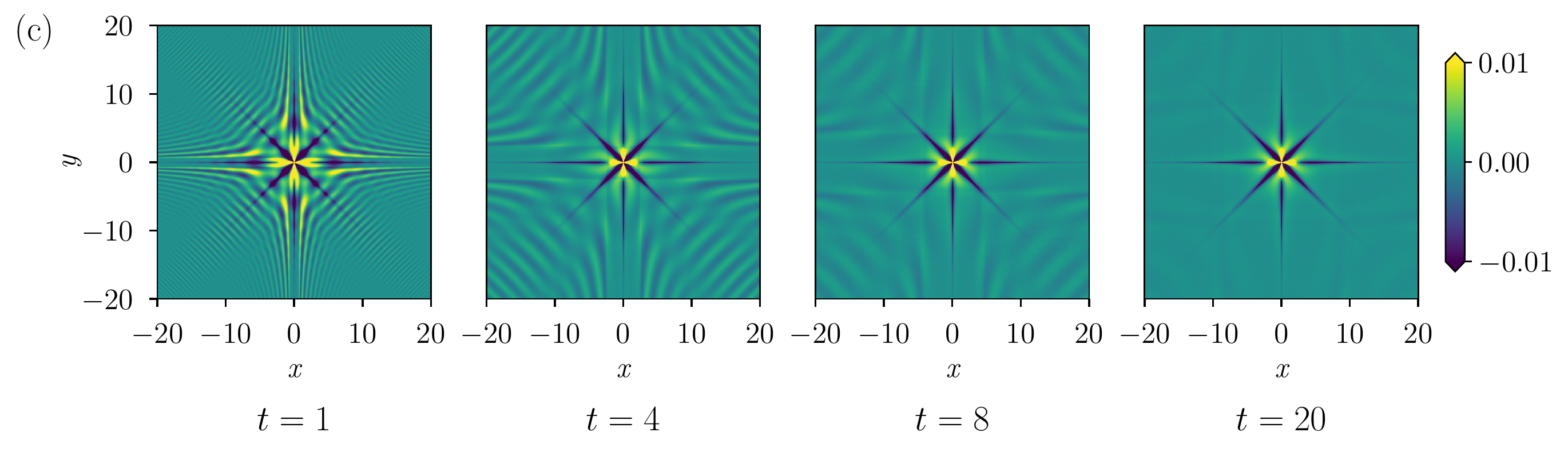}
    \caption{Regularized two-point correlation function~\eqref{eq:mass-quench} at different time moments when: \textit{(a)}~$\Z4$ symmetry is changed to the relativistic one, $m = 1$, $\eps = 0.1$; \textit{(b)}~$\Z4$-symmetric symmetry  is changed to $\Z8$; \textit{(c)}~$\Z8$ symmetry is changed to $\Z4$; for (b) and (c), $m = 1$, $\eps = 0.1$, $\delta = 0.01$.}
    \label{fig:DispChangeEvolPart2}
\end{figure}

Now let us consider how features corresponding to a particular symmetry choice before/after quench reveal themselves in the dynamics of the correlation function $G(t, x, y)$ given by~\eqref{eq:mass-quench}. To observe the spatial character of the dynamics, we study the spatial slices of $G(t, x, y)$ at fixed time moments (see \figref{fig:DispChangeEvolPart1} and \figref{fig:DispChangeEvolPart2}). Some difference is expected depending on whether, for example, the relativistic dispersion relation~\eqref{eq:reldisprel} is substituted for the regularized fractonic one~\eqref{eq:Z4disprel} and \eqref{eq:Z8disprel} or, vice versa, the fractonic dispersion relation is replaced by the relativistic one.

The evolution picture is highlighted by the following features:
\begin{itemize}
    \item Mass quench leaving the symmetries of the system unchanged leads to generation of slowly decaying waves. For the relativistic dispersion relation, the waves are spherically symmetric (as was studied before in~\cite{Calabrese:2007rg, Sotiriadis:2010si}), see \figref{fig:DispChangeEvolPart1}, row~(a).
    
    \item Mass quench in $\Z4$- or $\Z8$-symmetric fractonic theory induces two types of waves, as expected from the semiclassical analysis around eq.~\eqref{eq:vel}. At early times, the acausal waves fill the space outside the symmetry axes (i.e. along the ``cross'' for $\Z4$ symmetry or along the ``star'' for $\Z8$). After some time, one can observe the formation of a wave front in the form of expanding cross (or star), which is the wave front of the causal excitations emerging due to the relativistic regularization. The propagation of the causal excitations towards the region of acausal waves forms a mixed pattern, see \figref{fig:DispChangeEvolPart1}, row~(b). Quite similar behavior could be observed in the boundary type of quench (see \secref{sec:slab-quench} below), where we also consider the dependence on parameters $m$ and $\eps$. The fractonic nature of the quenched system becomes more elucidated in the finite volume theory with transformation of acausal waves into more complicated fractonic structures (see \secref{sec:finvolume}).
    
    \item The change of relativistic to fractonic symmetry produces a similar picture of evolution, see \figref{fig:DispChangeEvolPart1}, rows~(c) and (d). There is no imprint of the initial spherical symmetry that remains.
    
    \item Naively, one could expect that the general behavior for the fractonic to relativistic quench is similar to the relativistic to fractonic case, however, it differs drastically, see \figref{fig:DispChangeEvolPart2}, row~(a). The quench dynamics starts with the $\Z4$-symmetric wave and smoothly turns into the spherically symmetric one after some time, with the stationary correlations in the form of a cross being left as an imprint of the fractonic initial state.

    \item Switching the system symmetry from $\Z4$ to $\Z8$ leaves the causal waves propagating away from the symmetry axes --- the diagonals --- that are present after the quench but not before, see \figref{fig:DispChangeEvolPart2}, row~(b).
    
    \item Reverse switching from $\Z8$ to $\Z4$ symmetry induces a wave front along the final symmetry axes, see \figref{fig:DispChangeEvolPart2}, row~(c). However, one can observe that the system still bears an imprint of the initial symmetry at large times. The two-point function quickly gets to be stationary after the initial perturbation passes away from the origin.
\end{itemize}

\skipline
 
The results described above are obtained via straightforward application of formula~\eqref{eq:mass-quench} with some necessary modification. As we mentioned before, fractonic systems exhibit UV/IR mixing and high degeneracy of the ground state. To tame the divergences of the equilibrium system, we introduced the relativistic regulator~$\eps$ and the mass gap~$m$. One could expect that these regulators would be enough to calculate the non-equilibrium dynamics in a consistent way too. However, the integral kernel~\eqref{eq:FourierImage_mass} in the UV region shows the behavior responsible for the singular terms for some type of quenches. This highlights the subtleties of the fractonic systems getting more serious in the non-equilibrium situation.

The amplitude of the integral kernel $\mathcal{G}$~\eqref{eq:FourierImage_mass} for the purely relativistic mass quench in polar variables for large momenta has the form
\be
    \frac{\mathcal{G}(t, k, \varphi)}{\sin^2\om_k t}\Big|_{\text{rel.}\,\to\,\text{rel.}} \largemom \frac{m_0^2 - m^2}{2k^3} - \frac{(m_0^2 - m^2)(m_0^2 + 2m^2)}{4 k^5} + O\left(k^{-7}\right),
\ee
rapidly decaying for $k \to \infty$. However, for the quench of the type $\text{relativistic} \to \Z4$-symmetric we have for large $k$
\be
    \frac{\mathcal{G}(t, k, \varphi)}{\sin^2\om_k t}\Big|_{\text{rel.}\,\to\,\Z4} \largemom -\frac{1}{2k} + \frac{m^2 + 2(\sin\varphi)^{-2} + 2(\cos\varphi)^{-2}}{4 k^3} + O\left(k^{-5}\right).
\ee
which at the leading order contains singularity-generating term $\sim 1/k$, since its Fourier transform is $\sim 1/\sqrt{x^2 + y^2}$. In view of that, we regularize the integral kernel by subtraction of the term $-\sin^2\omega_k t/(2\sqrt{k_x^2 + k_y^2 + m^2})$, which removes undesirable features related to the divergences and singular behavior. The same holds for the quench of the type $\text{relativistic} \to \Z8$-symmetric. Changing the $\Z4$-symmetric dispersion relation to the relativistic one gives
\be
    \frac{\mathcal{G}(t, k, \varphi)}{\sin^2\om_k t}\Big|_{\Z4\,\to\,\text{rel.}} \largemom \frac{|\sin 2\varphi|}{4} + \frac{4\eps - 8 + m^2(\cos 4\varphi - 1)}{8|\sin 2\varphi|k^2} + O\left(k^{-4}\right),
\ee
which contains a constant term $c = |\sin 2\varphi|/4$ causing a singular behavior of the integral, thus, the term $c\,\sin^2\omega_k t$ is to be subtracted. The situation is more complicated with the asymptotics of~$\mathcal{G}$ in case of the $\Z8$ to $\Z4$ switch, since a singular-generating terms contribute along the diagonals $\varphi = \pm\pi/2$,
\be
    \frac{\mathcal{G}(t, k, \varphi)}{\sin^2\om_k t}\Big|_{\Z8\,\to\,\Z4} \underset{\substack{k\,\to\,\infty \\ \varphi\,\to\,\pm\pi/2}}{\approx} -\frac{1}{2\sqrt{\eps}k} + \frac{8\eps^2 + m^2}{4 \eps^{3/2}k^3} + O\left(k^{-4}\right),
\ee
or the $\Z4$ to $\Z8$ switch,
\be
    \frac{\mathcal{G}(t, k, \varphi)}{\sin^2\om_k t}\Big|_{\Z4\,\to\,\Z8} \underset{\substack{k\,\to\,\infty \\ \varphi\,\to\,\pm\pi/2}}{\approx} \frac{1}{4\eps} - \frac{2m^2/\eps^2 + 4}{8k^2} + O\left(k^{-4}\right).
\ee
Such large-$k$ behavior can be regularized by including a $\Z4$-symmetric part to the $\Z8$-symmetric dispersion relation,
\be
    \omega_k = \sqrt{\eps(k_x^2 + k_y^2) + \delta k_x^2 k_y^2 + \frac{4(k_x^3 k_y - k_x k_y^3)^2}{(k_x^2 + k_y^2)^2} + m^2}.
\ee
which turns the leading orders along the diagonals into $(\delta - 1)/(\sqrt{\delta}k^2)$ and $(1/\delta - 1)/k^2$, correspondingly, where $\delta$ is the regulating parameter.

\begin{figure}[t]\centering
    \includegraphics[width=0.38\textwidth]{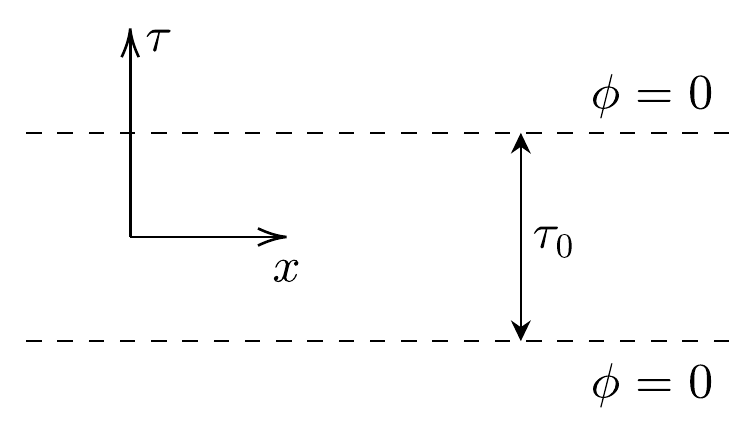}
	\caption{The path-integral geometry of the problem is a finite-size slab of width~$\tau_0$ with the Dirichlet boundary conditions imposed.}
    \label{fig:slab}
\end{figure}

\section{Boundary quenches}
\label{sec:slab-quench}

Now let us turn to the boundary quenches, another canonical model to study the out-of-equilibrium dynamics of field theory. In this type of quench, one effectively reduces the study of quench dynamics to the study of a theory which initial state is prepared on a finite-sized time slab with some boundary conditions imposed on the slab boundaries (here we choose the simplest case of the Dirichlet ones). In other words, the path-integral geometry of our system is a finite-size slab of width~$\tau_0$ (in Euclidean time~$\tau$) with the Dirichlet boundary conditions imposed on the slab boundaries (i.e. at $\tau = -\tau_0/2$ and $\tau = \tau_0/2$, see \figref{fig:slab}).

The real-time momentum-space two-point correlation function after the boundary quench is given by~\cite{Sotiriadis:2010si} 
\be
    \mathcal{F}(t_1, t_2, k_x, k_y) = \frac{\cos\om_k(t_1 - t_2)}{\om_k(e^{2\om_k \tau_0} - 1)} - \frac{e^{\om_k \tau_0}\cos\om_k(t_1 + t_2)}{\om_k(e^{2\om_k \tau_0} - 1)} + \frac{e^{-i\om_k|t_1 - t_2|}}{2\om_k}.
    \label{eq:FourierImage_slab}
\ee
In the position space, hence, the evolution of the equal-time two-point correlation function of field~$\phi$ with the correlation function at the $t = 0$ time moment subtracted is given by the Fourier transform
\be 
    \begin{aligned}
        G(t, x, y) & \equiv \corrfunc{\phi(t, x, y)\phi(t, 0, 0)} - \corrfunc{\phi(0, x, y)\phi(0, 0, 0)} = \\
        & = \int\frac{\,dk_x\,dk_y}{(2\pi)^2}e^{ik_xx + ik_yy}\frac{\sin^2(\om_k t)}{\om_k\sinh(\om_k \tau_0)}.
    \end{aligned}
    \label{eq:G_slab}
\ee
In this regularized two-point correlation function, we have a single dispersion relation encountered in the formula in contrast to the switching quench considered in \secref{sec:mass-quench}.

\begin{figure}[t]\centering
    \includegraphics[width=0.75\textwidth]{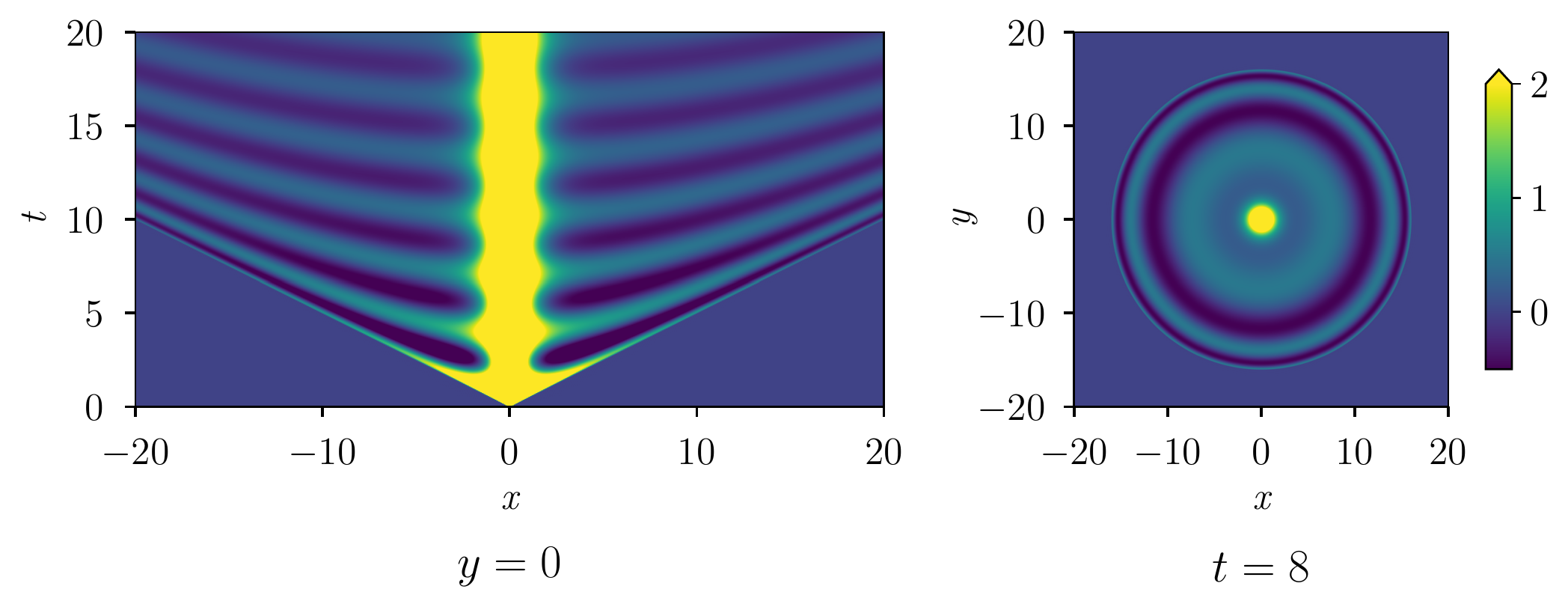}
    \caption{\textit{Left:}~the time dependence of the regularized two-point correlation function~\eqref{eq:G_slab} along the slice $y = 0$ in the boundary quench model with parameters $\tau_0 = 0.01$, $m = 1$. \textit{Right:}~the same at fixed time moment $t = 8$.}
    \label{fig:slab_rel}
\end{figure}

For the system in infinite volume, the qualitative evolution picture for boundary quenches is mostly the same as for mass-switching quench within one class of symmetries, see also \secref{sec:mass-quench}. Due to suppression by the~$\sinh$ factor in~\eqref{eq:G_slab}, no issue concerning potentially singular terms arises in contrast to the switching quench.

For the relativistic dispersion relation, we observe a spherically symmetric causally expanding perturbation (see \figref{fig:slab_rel}) as was obtained before in~\cite{Sotiriadis:2010si}. Now, let us consider what changes if we proceed in the same way with the fractonic dispersion relations, namely~\eqref{eq:Z4disprel} and \eqref{eq:Z8disprel}, which correspond to $\Z4$ and $\Z8$ choices of the correlation function symmetry, respectively.

\begin{figure}[t]\centering
    \includegraphics[width=0.9\textwidth]{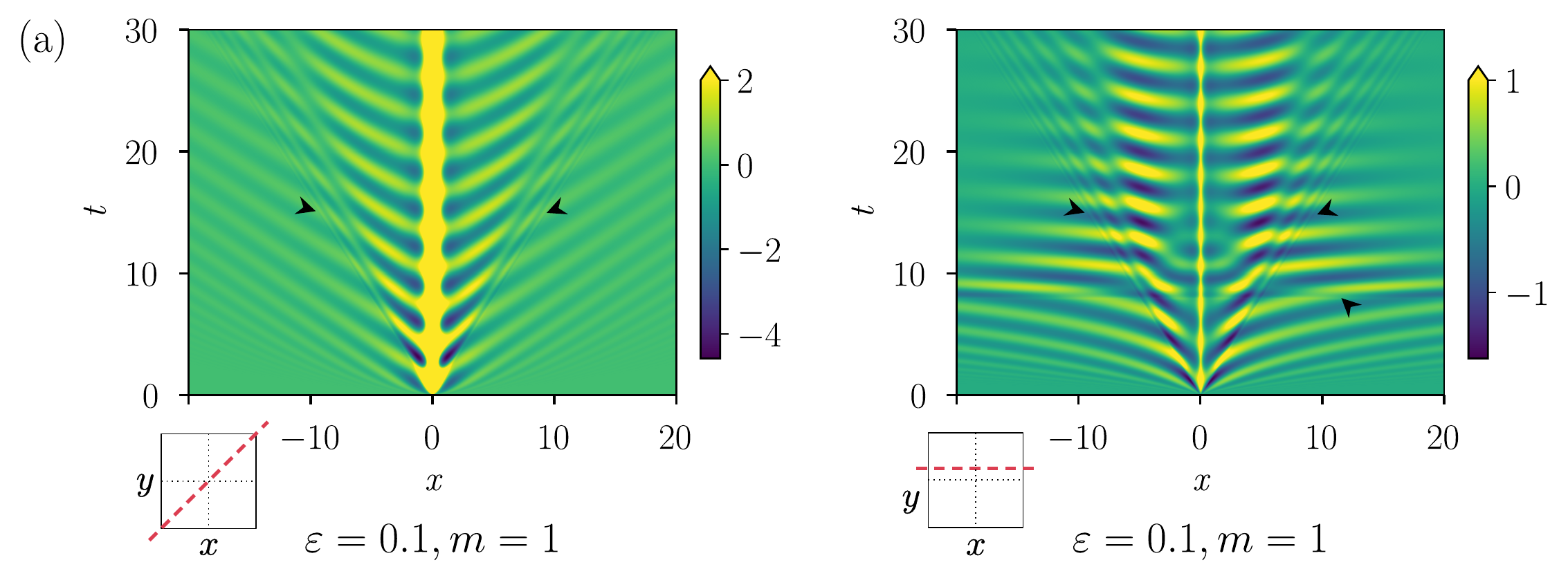}
    \includegraphics[width=0.9\textwidth]{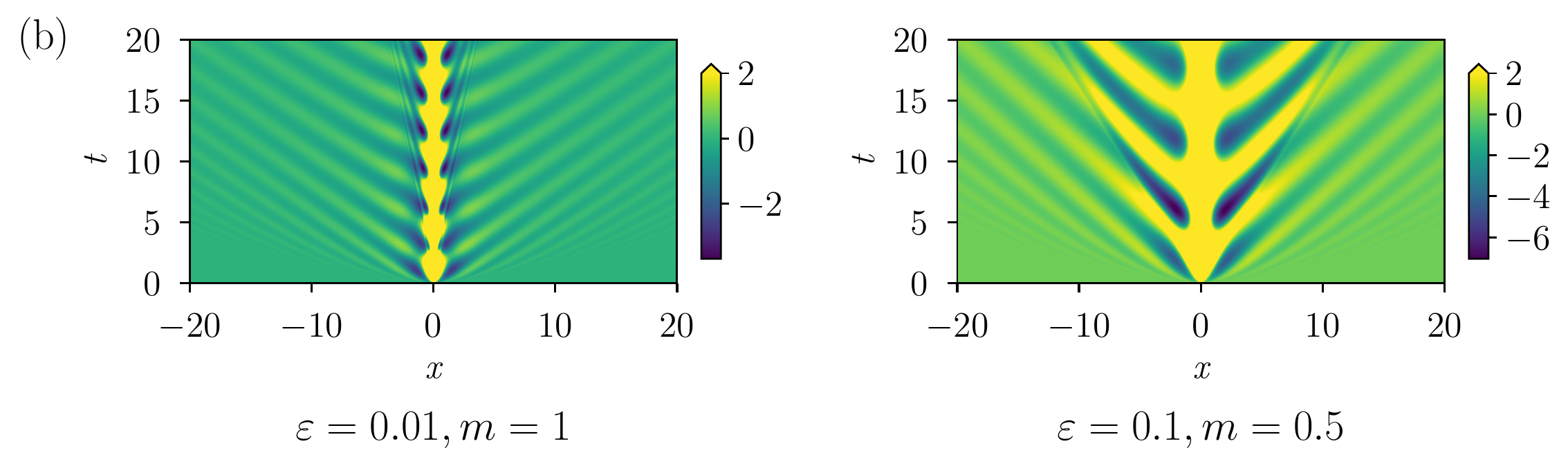}
	\caption{\textit{(a)}~Regularized two-point correlation function~\eqref{eq:G_slab} after boundary quench in the diagonal ($x = y$) slice of the spacetime (left) and in a horizontal ($y = 5$) slice (right). The parameters are $\tau_0 = 0.01$, $\eps = 0.1$, $m = 1$. Black arrowheads mark the wave fronts which distinguish two types of waves emanating from the origin. \textit{(b)}~The same for different values of~$\eps$ and~$m$, $\eps = 0.01$, $m = 1$ (left), $\eps = 0.1$, $m = 0.5$ (right). For both figures, $\tau_0 = 0.01$.}
    \label{fig:Z4_xt}
\end{figure}

\begin{figure}[t]\centering
    \includegraphics[width=1.\textwidth]{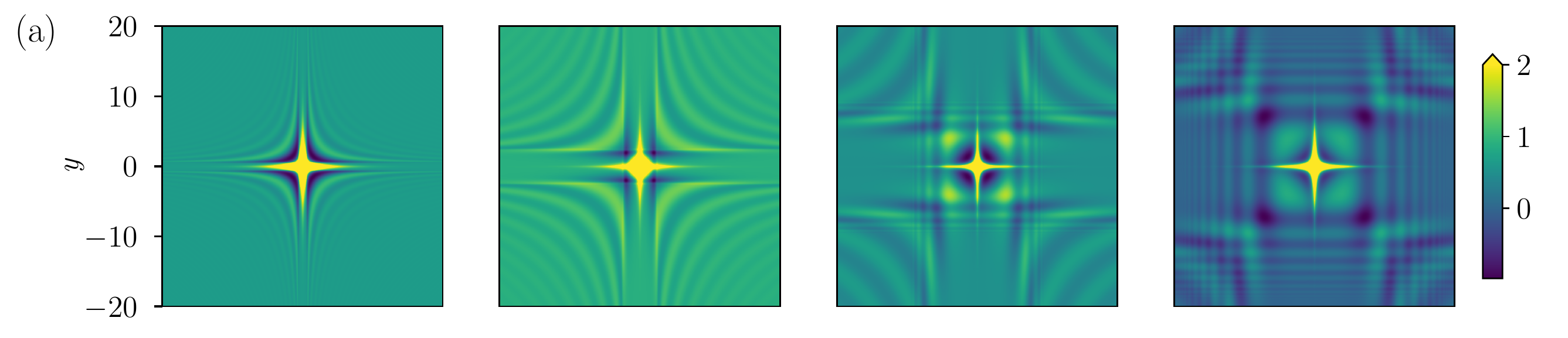}
    \includegraphics[width=1.\textwidth]{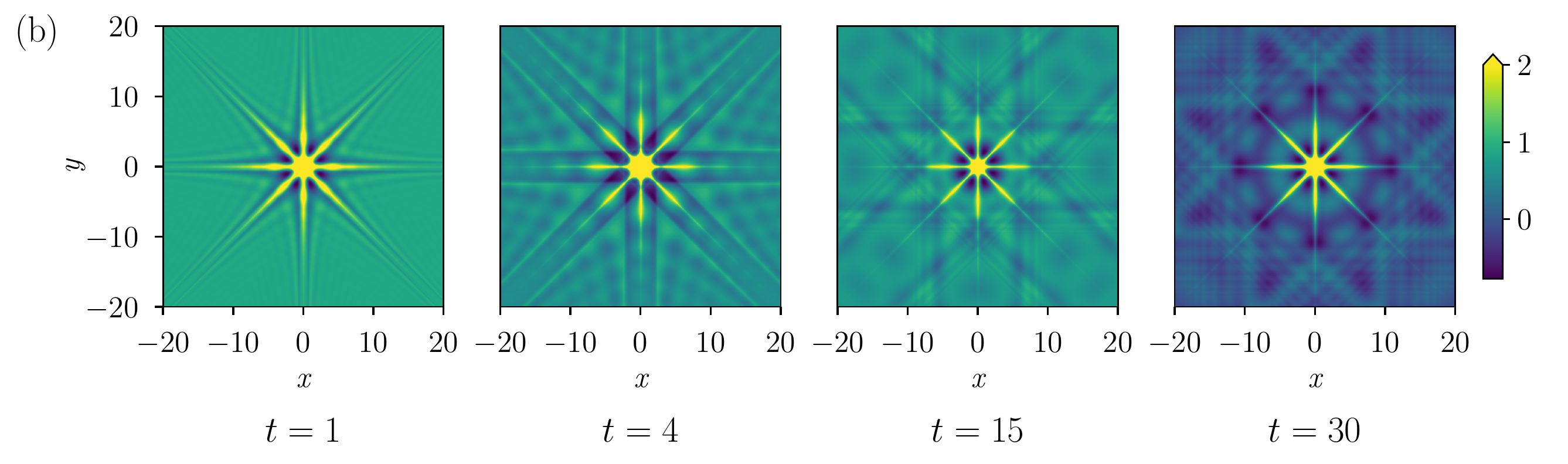}
    \caption{\textit{(a)}~Spatial distribution of perturbations~\eqref{eq:G_slab} in the $\Z4$-symmetric model of boundary quench for fixed time moments. There are two domains: in the inner one, there are relativistic-like waves, while in the outer one, the hyperbolic-shaped waves. At large enough times, see for example the plot for $t = 30$, the amplitude of the wave front distinguishing these domains becomes smaller and two types of waves get mixed. \textit{(b)}~The same for the $\Z8$-symmetric model. For both rows, the parameters are $\tau_0 = 0.01$, $\eps = 0.1$, $m = 1$.}
    \label{fig:SlabEvol}
\end{figure}

\begin{itemize}
    \item We observe that the time-dependence of the two-point function is described by two types of $\Z4$-symmetric waves propagating with different velocity; governed by the relativistic regularization (``relativistic waves'') inside an effective light cone and the other, hyperbolic-shaped, type of waves (``fractonic waves'') starting, in turn, outside this light cone. The front of the ``relativistic waves'' has the form of a domain wall: a straight line stretching through the space along $x$ and $y$ axes. For large times, the wave front amplitude falls and the two types of waves get mixed. In \figref{fig:Z4_xt}, row~(a), we highlight the time and spatial dependence of the correlation function for different orientations of the reference plane. In \figref{fig:SlabEvol}, row~(a), we plot the spatial dependence of the correlator at fixed time moments.
     
    \item The dependence on the relativistic regulator and the mass can be seen in \figref{fig:Z4_xt}, row~(b). Since~$\eps$ corresponds to the maximum velocity of the relativistic-like perturbations, making it larger squeezes the cone formed by the effective horizon. In the pure massive fractonic limit, $\eps \to 0$, it is expected that only hyperbolic-shaped waves should remain since the effective velocity of the relativistic-like waves tends to zero. Making mass smaller leads to zooming in the region about the origin with amplifying the amplitude.

    \item In the case of the $\Z8$-symmetric model, the picture does not change qualitatively, see spatial slices in \figref{fig:SlabEvol}, row~(b), but the symmetry of correlations changes to $\Z8$ as should be expected. There are still two types of propagating waves with a separating them wave front that vanishes for large enough times.
\end{itemize}

\section{Global quenches in finite volume theory}
\label{sec:finvolume}

\begin{figure}[t]\centering
    \includegraphics[width=1.\textwidth]{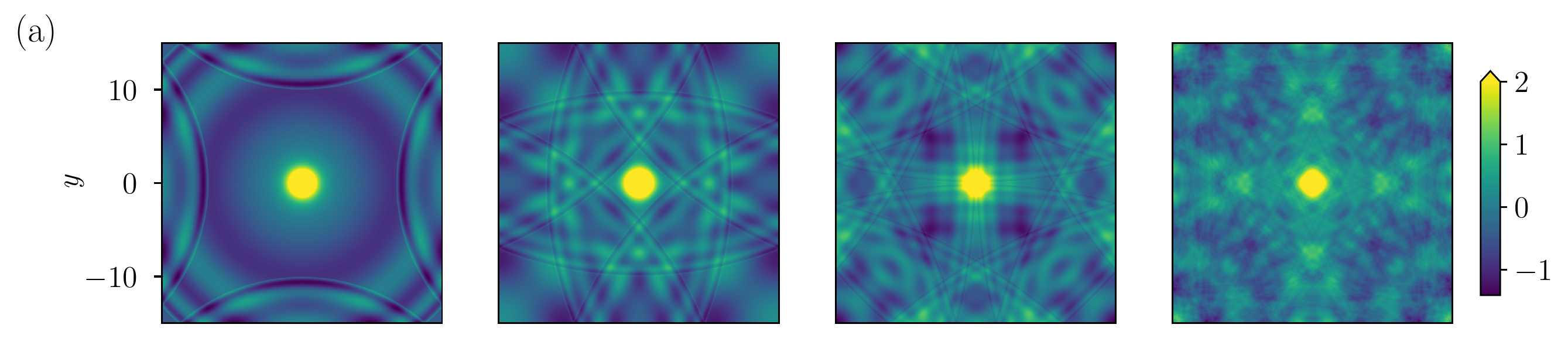}
	\includegraphics[width=1.\textwidth]{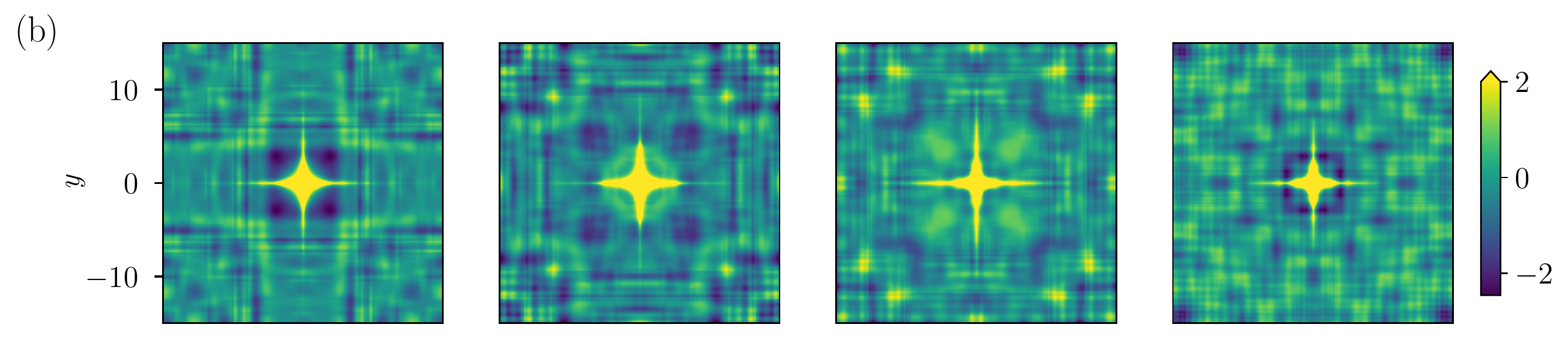}
	\includegraphics[width=1.\textwidth]{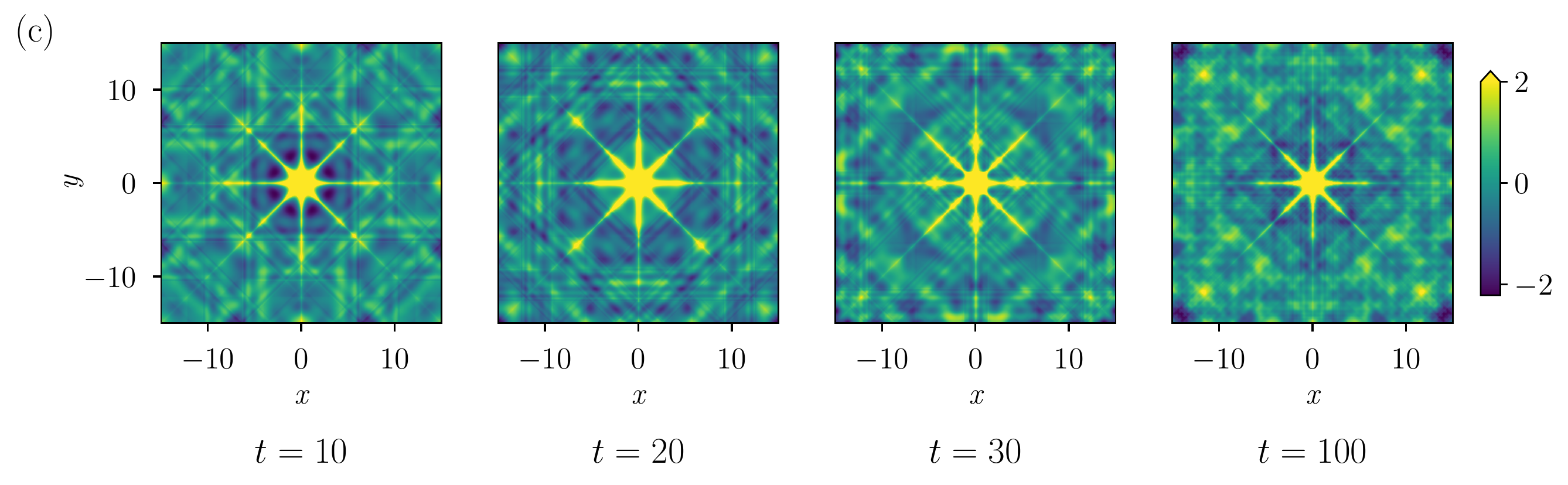}
	\caption{\textit{(a)}~Spatial distribution of perturbations~\eqref{eq:G_slab_finvol} in the relativistic finite volume model of boundary quench for fixed time moments. The parameters are $L_x = L_y = 30$, $\tau_0 = 0.01$, $m = 1$. Several windings round the spacetime form a folded discrete $\Z4$-symmetric wave front. \textit{(b)}~The same for the $\Z4$-symmetric finite volume model for fixed time moments. One can compare plots for small times, $t < L_x$, $t < L_y$, with the infinite-volume case, \figref{fig:SlabEvol}: there is the contribution of the waves of high frequency. \textit{(c)}~The same for the $\Z8$-symmetric finite volume model. For~(b) and~(c), the parameters are $L_x = L_y = 30$, $\tau_0 = 0.01$, $\eps = 0.1$, $m = 1$.}
	\label{fig:FinVolEvol}
\end{figure}

Now let us consider a torus with periods $L_x$ and $L_y$, i.e., $x \sim x + L_x$, $y \sim y + L_y$ and study how the non-equilibrium theory after the boundary quench in finite volume differs from its $\mathbb{R}^2$-counterpart. Imposing the periodic boundary conditions with respect to the space coordinates implies the periodicity of the two-point function, $G(t, x, y) = G(t, x + L_x, y) = G(t, x, y + L_y)$. The corresponding solution can be built up from the same Fourier image as for the infinite spacetime case~\eqref{eq:FourierImage_slab} but with the substitution of the continuous Fourier transform for its discrete analogue,
\be
    k_x \to k_n, \quad \int\frac{dk_x}{2\pi} \to \frac{1}{L_x}\sum_n,
\ee
where $k_n = 2\pi n/L_x$ is the discrete frequencies. Hence, we get the regularized two-point function defined as
\be
    \begin{aligned}
        G(t, x, y) & \equiv \corrfunc{\phi(t, x, y)\phi(t, 0, 0)} - \corrfunc{\phi(0, x, y)\phi(0, 0, 0)} = \\ 
        & = \frac{1}{L_x L_y}\sum_{n\,= -\infty}^{\infty}\sum_{s\,= -\infty}^{\infty} e^{ik_n x + ik_s y}\frac{\sin^2(\om_{ns} t)}{\om_{ns}\sinh(\om_{ns}\tau_0)},
    \end{aligned}
    \label{eq:G_slab_finvol}
\ee
where $\om_{ns}$ is the discrete analogue of the dispersion relation which for the relativistic theory, the $\Z4$-symmetric theory and the $\Z8$-symmetric theory, correspondingly, is
\begin{gather}
    \om_{ns} = \sqrt{k_n^2 + k_s^2 + m^2}, \\
    \om_{ns} = \sqrt{\eps(k_n^2 + k_s^2) + k_n^2 k_s^2 + m^2}, \\
    \om_{ns} = \sqrt{\eps(k_n^2 + k_s^2) + \frac{4\left(k_n^3 k_s - k_n k_s^3\right)^2}{\left(k_n^2 + k_s^2\right)^2} + m^2}.
\end{gather}

\skipline

\begin{figure}[t]\centering
    \includegraphics[width=1.\textwidth]{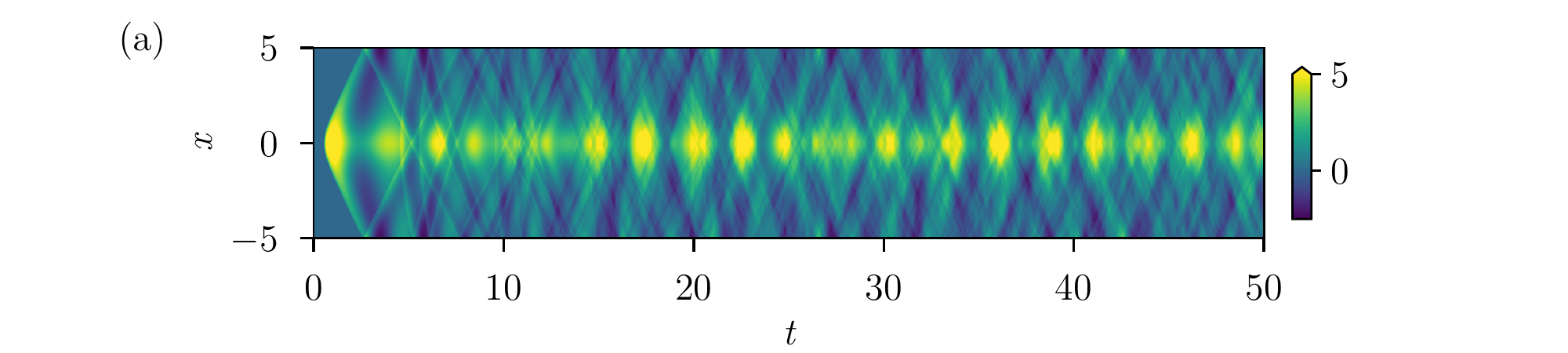}
    \includegraphics[width=1.\textwidth]{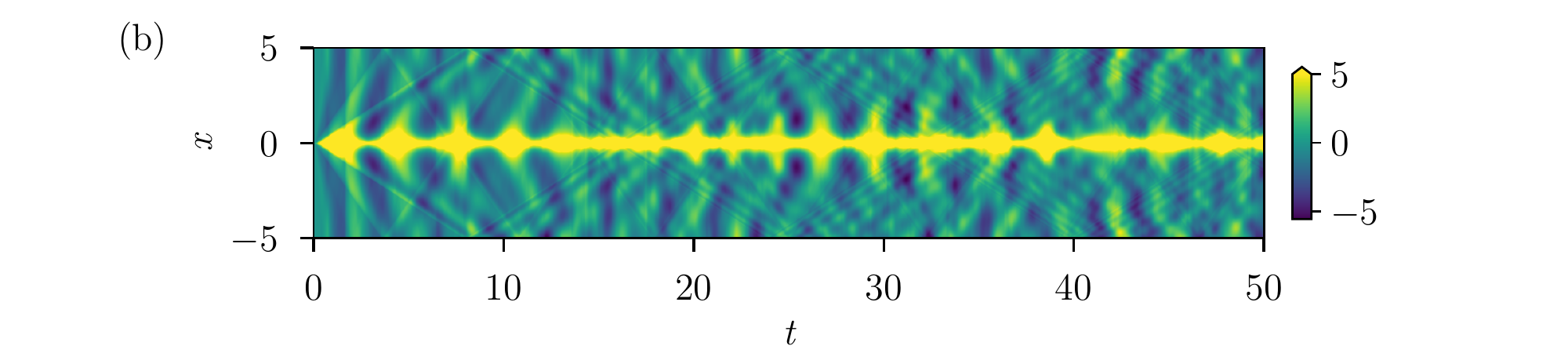}
    \includegraphics[width=1.\textwidth]{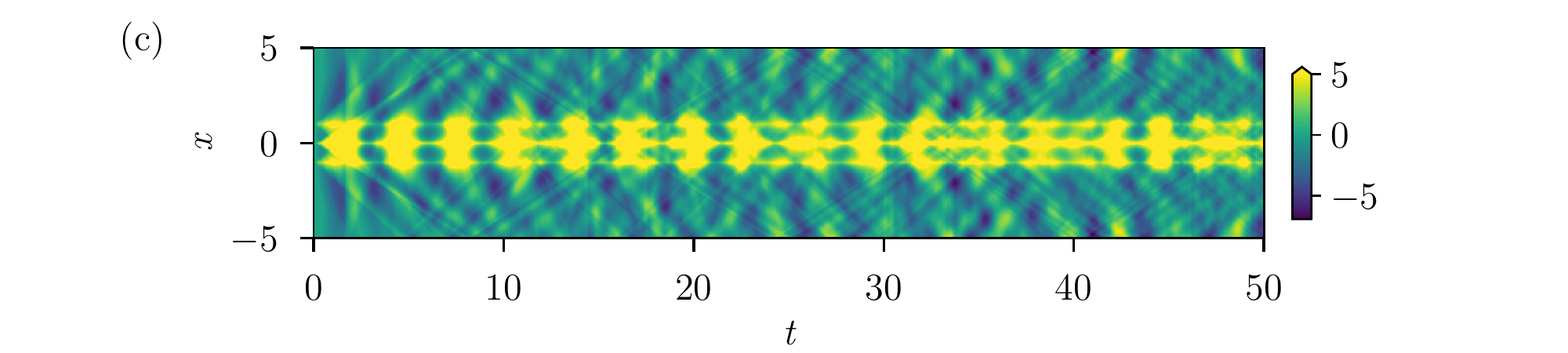}
    \caption{Regularized two-point correlation function after boundary quench in a horizontal ($y = 1$) slice for the: \textit{(a)}~relativistic finite volume model, \textit{(b)}~$\Z4$-symmetric finite volume model and \textit{(c)}~$\Z8$-symmetric finite volume model. The parameters are $L_x = L_y = 10$, $\tau_0 = 0.01$, $m = 1$ for all cases and $\eps = 0.1$ for the fractonic cases.}
    \label{fig:FinVol_xt}
\end{figure}

In the relativistic case (see \figref{fig:FinVolEvol}, row~(a) and \figref{fig:FinVol_xt}, row~(a)), initially spherically symmetric wave front winds around spacetime after reaching the boundaries. Several windings propagating in different directions form complicated but still highly-regular picture of folded wave front. Imposing of boundaries leads to break of the continuous rotational symmetry with $\Z4$ symmetry remaining.

The cases corresponding to $\Z4$ and $\Z8$ symmetry (see \figref{fig:FinVolEvol}, rows~(b) and~(c)) differ drastically from the relativistic one. High-frequency acausal patterns made up of ``lines'', filling all the space, emerges already at small (in comparison with the period of spacetime, $t < L_x$, $t < L_y$) times. This effect is due to the already mentioned hyperbolic-shaped fractonic waves which have non-vanishing amplitude at the boundaries. After the initial wave front reaches the boundary and several windings occur, a randomly looking pattern of localization-delocalization forms with the mean density keeping approximately uniform. The slice along one spatial coordinate for $\Z4$-symmetric case, \figref{fig:FinVol_xt}, row~(b), reveals an additional fractonic feature. We have discussed that a wave front emerges from the origin due to the relativistic regularization. In finite volume, one can note the secondary and even higher order wave fronts that originate immediately after the quench.

\section{Conclusions}
\label{sec:conclusions}

In this paper, we have studied dynamics of fractonic quantum field systems taken out of equilibrium by global quantum quenches. Field theories under consideration exhibit exotic symmetries of fractonic phases of matter, namely, $\Z{n}$ discrete rotational symmetry being opposite to Lorentz rotational symmetry of the relativistic theory.

There are several quench setups that can serve as the probe of the system dynamics:
\begin{itemize}
    \item instant change of a mass gap with the symmetry of the system left fixed;
    \item instant change of a symmetry (i.e. turning on and off restricted mobility in the system);
    \item boundary quench.
\end{itemize}

We have observed that propagation of fractonic perturbations after the mass quench or the boundary quench occurs acausally in the regions between the axes of symmetry. The second setup of switching symmetry is less predictable. The evolution picture depends strongly on the order in which the type of symmetry is changed. In fact, switching between relativistic and fractonic cases produces a picture similar to quench by mass in fractonic theory with no sign of the initial Lorentz symmetry remaining, while the reversed order change clearly demonstrates an imprint of the initial $\Z4$ symmetry.

The point of view according to which the form of the dispersion relation determines quench dynamics turns to be useful for generalizations. Indeed, a simple change of the polar variables period of an invariant polynomial allows the fractonic $\Z4$ symmetry to be transformed to higher orders, with $\Z8$ symmetry as a particular example that we have considered. From a technical point of view, fractonic theories require regularization of divergences which follow from the ground-state degeneracy. We have introduced a regularization by adding a small amount of relativistic degrees of freedom in the fractonic theory which can be taken into account by adding the Lorentz invariant polynomial into the dispersion relation. This is enough to eliminate divergences of the propagator or the non-equilibrium two-point correlation function in the boundary quench. However, the case of switching from or to fractonic theory requires additional subtraction of terms that generate singularities.

Finally, we have found that the feature of restricted mobility of the fractonic field theory is especially clear in finite volume. In this case, the domain walls, stretching along the symmetry axes, are turning into $\Z{n}$-symmetric localized patterns after some short period of evolution.

\acknowledgments

We would like to thank A. I. Belokon for careful reading and useful comments on the manuscript.

\bibliography{main}

\providecommand{\href}[2]{#2}\begingroup\raggedright\begin{thebibliography}{10}

\bibitem{Calabrese:2006rx}
P.~Calabrese and J.L.~Cardy, \emph{{Time-dependence of correlation functions
  following a quantum quench}},
  \href{https://doi.org/10.1103/PhysRevLett.96.136801}{\emph{Phys. Rev. Lett.}
  {\bfseries 96} (2006) 136801}
  [\href{https://arxiv.org/abs/cond-mat/0601225}{{\ttfamily
  cond-mat/0601225}}].

\bibitem{Calabrese:2007rg}
P.~Calabrese and J.~Cardy, \emph{{Quantum Quenches in Extended Systems}},
  \href{https://doi.org/10.1088/1742-5468/2007/06/P06008}{\emph{J. Stat. Mech.}
  {\bfseries 0706} (2007) P06008}
  [\href{https://arxiv.org/abs/0704.1880}{{\ttfamily 0704.1880}}].

\bibitem{Das:2014hqa}
S.R.~Das, D.A.~Galante and R.C.~Myers, \emph{{Universality in fast quantum
  quenches}}, \href{https://doi.org/10.1007/JHEP02(2015)167}{\emph{JHEP}
  {\bfseries 02} (2015) 167} [\href{https://arxiv.org/abs/1411.7710}{{\ttfamily
  1411.7710}}].

\bibitem{Calabrese:2016xau}
P.~Calabrese and J.~Cardy, \emph{{Quantum quenches in 1 + 1 dimensional
  conformal field theories}},
  \href{https://doi.org/10.1088/1742-5468/2016/06/064003}{\emph{J. Stat. Mech.}
  {\bfseries 1606} (2016) 064003}
  [\href{https://arxiv.org/abs/1603.02889}{{\ttfamily 1603.02889}}].

\bibitem{Sotiriadis2009thermal}
S.~Sotiriadis, P.~Calabrese and J.~Cardy, \emph{Quantum quench from a thermal
  initial state}, \href{https://doi.org/10.1209/0295-5075/87/20002}{\emph{EPL
  (Europhysics Letters)} {\bfseries 87} (2009) 20002}
  [\href{https://arxiv.org/abs/0903.0895}{{\ttfamily 0903.0895}}].

\bibitem{Rajabpour:2014osa}
M.A.~Rajabpour and S.~Sotiriadis, \emph{{Quantum quench in long-range field
  theories}}, \href{https://doi.org/10.1103/PhysRevB.91.045131}{\emph{Phys.
  Rev. B} {\bfseries 91} (2015) 045131}
  [\href{https://arxiv.org/abs/1409.6558}{{\ttfamily 1409.6558}}].

\bibitem{Das:2015jka}
S.R.~Das, D.A.~Galante and R.C.~Myers, \emph{{Smooth and fast versus
  instantaneous quenches in quantum field theory}},
  \href{https://doi.org/10.1007/JHEP08(2015)073}{\emph{JHEP} {\bfseries 08}
  (2015) 073} [\href{https://arxiv.org/abs/1505.05224}{{\ttfamily
  1505.05224}}].

\bibitem{Das:2016lla}
S.R.~Das, D.A.~Galante and R.C.~Myers, \emph{{Quantum Quenches in Free Field
  Theory: Universal Scaling at Any Rate}},
  \href{https://doi.org/10.1007/JHEP05(2016)164}{\emph{JHEP} {\bfseries 05}
  (2016) 164} [\href{https://arxiv.org/abs/1602.08547}{{\ttfamily
  1602.08547}}].

\bibitem{Sotiriadis:2010si}
S.~Sotiriadis and J.~Cardy, \emph{{Quantum quench in interacting field theory:
  A Self-consistent approximation}},
  \href{https://doi.org/10.1103/PhysRevB.81.134305}{\emph{Phys. Rev. B}
  {\bfseries 81} (2010) 134305}
  [\href{https://arxiv.org/abs/1002.0167}{{\ttfamily 1002.0167}}].

\bibitem{He:2019vzf}
S.~He and H.~Shu, \emph{{Correlation functions, entanglement and chaos in the $
  T\overline{T}/J\overline{T} $-deformed CFTs}},
  \href{https://doi.org/10.1007/JHEP02(2020)088}{\emph{JHEP} {\bfseries 02}
  (2020) 088} [\href{https://arxiv.org/abs/1907.12603}{{\ttfamily
  1907.12603}}].

\bibitem{Ageev:2022kpm}
D.S.~Ageev, A.I.~Belokon and V.V.~Pushkarev, \emph{{From locality to
  irregularity: introducing local quenches in massive scalar field theory}},
  \href{https://doi.org/10.1007/JHEP05(2023)188}{\emph{JHEP} {\bfseries 05}
  (2023) 188} [\href{https://arxiv.org/abs/2205.12290}{{\ttfamily
  2205.12290}}].

\bibitem{Danielsson:1999fa}
U.H.~Danielsson, E.~Keski-Vakkuri and M.~Kruczenski, \emph{{Black hole
  formation in AdS and thermalization on the boundary}},
  \href{https://doi.org/10.1088/1126-6708/2000/02/039}{\emph{JHEP} {\bfseries
  02} (2000) 039} [\href{https://arxiv.org/abs/hep-th/9912209}{{\ttfamily
  hep-th/9912209}}].

\bibitem{Balasubramanian:2010ce}
V.~Balasubramanian, A.~Bernamonti, J.~de~Boer, N.~Copland, B.~Craps,
  E.~Keski-Vakkuri et~al., \emph{{Thermalization of Strongly Coupled Field
  Theories}}, \href{https://doi.org/10.1103/PhysRevLett.106.191601}{\emph{Phys.
  Rev. Lett.} {\bfseries 106} (2011) 191601}
  [\href{https://arxiv.org/abs/1012.4753}{{\ttfamily 1012.4753}}].

\bibitem{Balasubramanian:2011ur}
V.~Balasubramanian, A.~Bernamonti, J.~de~Boer, N.~Copland, B.~Craps,
  E.~Keski-Vakkuri et~al., \emph{{Holographic Thermalization}},
  \href{https://doi.org/10.1103/PhysRevD.84.026010}{\emph{Phys. Rev. D}
  {\bfseries 84} (2011) 026010}
  [\href{https://arxiv.org/abs/1103.2683}{{\ttfamily 1103.2683}}].

\bibitem{Buchel:2013gba}
A.~Buchel, R.C.~Myers and A.~van Niekerk, \emph{{Universality of Abrupt
  Holographic Quenches}},
  \href{https://doi.org/10.1103/PhysRevLett.111.201602}{\emph{Phys. Rev. Lett.}
  {\bfseries 111} (2013) 201602}
  [\href{https://arxiv.org/abs/1307.4740}{{\ttfamily 1307.4740}}].

\bibitem{Asplund:2014coa}
C.T.~Asplund, A.~Bernamonti, F.~Galli and T.~Hartman, \emph{{Holographic
  Entanglement Entropy from 2d CFT: Heavy States and Local Quenches}},
  \href{https://doi.org/10.1007/JHEP02(2015)171}{\emph{JHEP} {\bfseries 02}
  (2015) 171} [\href{https://arxiv.org/abs/1410.1392}{{\ttfamily 1410.1392}}].

\bibitem{Ageev:2017oku}
D.S.~Ageev and I.Y.~Aref'eva, \emph{{Waking and scrambling in holographic
  heating up}}, \href{https://doi.org/10.1134/S0040577917100105}{\emph{Teor.
  Mat. Fiz.} {\bfseries 193} (2017) 146}
  [\href{https://arxiv.org/abs/1701.07280}{{\ttfamily 1701.07280}}].

\bibitem{Ageev:2017wet}
D.S.~Ageev and I.Y.~Aref'eva, \emph{{Holographic Non-equilibrium Heating}},
  \href{https://doi.org/10.1007/JHEP03(2018)103}{\emph{JHEP} {\bfseries 03}
  (2018) 103} [\href{https://arxiv.org/abs/1704.07747}{{\ttfamily
  1704.07747}}].

\bibitem{Calabrese:2005in}
P.~Calabrese and J.L.~Cardy, \emph{{Evolution of entanglement entropy in
  one-dimensional systems}},
  \href{https://doi.org/10.1088/1742-5468/2005/04/P04010}{\emph{J. Stat. Mech.}
  {\bfseries 0504} (2005) P04010}
  [\href{https://arxiv.org/abs/cond-mat/0503393}{{\ttfamily
  cond-mat/0503393}}].

\bibitem{Nozaki:2014hna}
M.~Nozaki, T.~Numasawa and T.~Takayanagi, \emph{{Quantum Entanglement of Local
  Operators in Conformal Field Theories}},
  \href{https://doi.org/10.1103/PhysRevLett.112.111602}{\emph{Phys. Rev. Lett.}
  {\bfseries 112} (2014) 111602}
  [\href{https://arxiv.org/abs/1401.0539}{{\ttfamily 1401.0539}}].

\bibitem{He:2014mwa}
S.~He, T.~Numasawa, T.~Takayanagi and K.~Watanabe, \emph{{Quantum dimension as
  entanglement entropy in two dimensional conformal field theories}},
  \href{https://doi.org/10.1103/PhysRevD.90.041701}{\emph{Phys. Rev. D}
  {\bfseries 90} (2014) 041701}
  [\href{https://arxiv.org/abs/1403.0702}{{\ttfamily 1403.0702}}].

\bibitem{Caputa:2014eta}
P.~Caputa, J.~Sim\'on, A.~\v{S}tikonas and T.~Takayanagi, \emph{{Quantum
  Entanglement of Localized Excited States at Finite Temperature}},
  \href{https://doi.org/10.1007/JHEP01(2015)102}{\emph{JHEP} {\bfseries 01}
  (2015) 102} [\href{https://arxiv.org/abs/1410.2287}{{\ttfamily 1410.2287}}].

\bibitem{Cotler:2016acd}
J.S.~Cotler, M.P.~Hertzberg, M.~Mezei and M.T.~Mueller, \emph{{Entanglement
  Growth after a Global Quench in Free Scalar Field Theory}},
  \href{https://doi.org/10.1007/JHEP11(2016)166}{\emph{JHEP} {\bfseries 11}
  (2016) 166} [\href{https://arxiv.org/abs/1609.00872}{{\ttfamily
  1609.00872}}].

\bibitem{Mozaffar:2021nex}
M.R.M.~Mozaffar and A.~Mollabashi, \emph{{Time scaling of entanglement in
  integrable scale-invariant theories}},
  \href{https://doi.org/10.1103/PhysRevResearch.4.L022010}{\emph{Phys. Rev.
  Res.} {\bfseries 4} (2022) L022010}
  [\href{https://arxiv.org/abs/2106.14700}{{\ttfamily 2106.14700}}].

\bibitem{MohammadiMozaffar:2018vmk}
M.R.~Mohammadi~Mozaffar and A.~Mollabashi, \emph{{Entanglement Evolution in
  Lifshitz-type Scalar Theories}},
  \href{https://doi.org/10.1007/JHEP01(2019)137}{\emph{JHEP} {\bfseries 01}
  (2019) 137} [\href{https://arxiv.org/abs/1811.11470}{{\ttfamily
  1811.11470}}].

\bibitem{MohammadiMozaffar:2019gpn}
M.R.~Mohammadi~Mozaffar and A.~Mollabashi, \emph{{Universal Scaling in Fast
  Quenches Near Lifshitz-Like Fixed Points}},
  \href{https://doi.org/10.1016/j.physletb.2019.134906}{\emph{Phys. Lett. B}
  {\bfseries 797} (2019) 134906}
  [\href{https://arxiv.org/abs/1906.07017}{{\ttfamily 1906.07017}}].

\bibitem{Vijay:2015mka}
S.~Vijay, J.~Haah and L.~Fu, \emph{{A New Kind of Topological Quantum Order: A
  Dimensional Hierarchy of Quasiparticles Built from Stationary Excitations}},
  \href{https://doi.org/10.1103/PhysRevB.92.235136}{\emph{Phys. Rev. B}
  {\bfseries 92} (2015) 235136}
  [\href{https://arxiv.org/abs/1505.02576}{{\ttfamily 1505.02576}}].

\bibitem{Pretko:2018jbi}
M.~Pretko, \emph{{The Fracton Gauge Principle}},
  \href{https://doi.org/10.1103/PhysRevB.98.115134}{\emph{Phys. Rev. B}
  {\bfseries 98} (2018) 115134}
  [\href{https://arxiv.org/abs/1807.11479}{{\ttfamily 1807.11479}}].

\bibitem{Gromov2019towards}
A.~Gromov, \emph{Towards classification of fracton phases: the multipole
  algebra}, \href{https://doi.org/10.1103/PhysRevX.9.031035}{\emph{Physical
  Review X} {\bfseries 9} (2019) 031035}.

\bibitem{Radzihovsky:2019jdo}
L.~Radzihovsky and M.~Hermele, \emph{{Fractons from vector gauge theory}},
  \href{https://doi.org/10.1103/PhysRevLett.124.050402}{\emph{Phys. Rev. Lett.}
  {\bfseries 124} (2020) 050402}
  [\href{https://arxiv.org/abs/1905.06951}{{\ttfamily 1905.06951}}].

\bibitem{Seiberg:2019vrp}
N.~Seiberg, \emph{{Field Theories With a Vector Global Symmetry}},
  \href{https://doi.org/10.21468/SciPostPhys.8.4.050}{\emph{SciPost Phys.}
  {\bfseries 8} (2020) 050} [\href{https://arxiv.org/abs/1909.10544}{{\ttfamily
  1909.10544}}].

\bibitem{Seiberg:2020bhn}
N.~Seiberg and S.-H.~Shao, \emph{{Exotic Symmetries, Duality, and Fractons in
  2+1-Dimensional Quantum Field Theory}},
  \href{https://doi.org/10.21468/SciPostPhys.10.2.027}{\emph{SciPost Phys.}
  {\bfseries 10} (2021) 027}
  [\href{https://arxiv.org/abs/2003.10466}{{\ttfamily 2003.10466}}].

\bibitem{Seiberg:2020wsg}
N.~Seiberg and S.-H.~Shao, \emph{{Exotic $U(1)$ Symmetries, Duality, and
  Fractons in 3+1-Dimensional Quantum Field Theory}},
  \href{https://doi.org/10.21468/SciPostPhys.9.4.046}{\emph{SciPost Phys.}
  {\bfseries 9} (2020) 046} [\href{https://arxiv.org/abs/2004.00015}{{\ttfamily
  2004.00015}}].

\bibitem{Distler:2021bop}
J.~Distler, J.~Distler, M.~Jafry, M.~Jafry, A.~Karch, A.~Karch et~al.,
  \emph{{Interacting fractons in 2+1-dimensional quantum field theory}},
  \href{https://doi.org/10.1007/JHEP03(2022)070}{\emph{JHEP} {\bfseries 03}
  (2022) 070} [\href{https://arxiv.org/abs/2112.05726}{{\ttfamily
  2112.05726}}].

\bibitem{Bulmash:2023msp}
D.~Bulmash, O.~Hart and R.~Nandkishore, \emph{{Multipole groups and fracton
  phenomena on arbitrary crystalline lattices}},
  \href{https://arxiv.org/abs/2301.10782}{{\ttfamily 2301.10782}}.

\bibitem{Khlopov1981fractionally}
M.Y.~Khlopov, \emph{Fractionally charged particles and quark confinement},
  {\emph{JETP Lett.(Engl. Transl.)} {\bfseries 33} (1981) }.

\bibitem{Paramekanti2002ring}
A.~Paramekanti, L.~Balents and M.P.A.~Fisher, \emph{Ring exchange, the exciton
  bose liquid, and bosonization in two dimensions},
  \href{https://doi.org/10.1103/PhysRevB.66.054526}{\emph{Phys. Rev. B}
  {\bfseries 66} (2002) 054526}
  [\href{https://arxiv.org/abs/cond-mat/0203171}{{\ttfamily
  cond-mat/0203171}}].

\bibitem{Islam:2023cmm}
M.M.~Islam, K.~Sengupta and R.~Sensarma, \emph{{Non-equilibrium dynamics of
  bosons with dipole symmetry: A large $N$ Keldysh approach}},
  \href{https://arxiv.org/abs/2305.13372}{{\ttfamily 2305.13372}}.

\end{thebibliography}\endgroup
\bibliographystyle{JHEP}

\end{document}